\documentstyle[aps,multicol,prl]{revtex}
\renewcommand{\narrowtext}{\begin{multicols}{2} \global\columnwidth20.5pc}
\renewcommand{\widetext}{\end{multicols} \global\columnwidth42.5pc}
\input{epsf.tex}
\def\inseps#1#2{\def\epsfsize##1##2{#2##1} \centerline{\epsfbox{#1}}}

\begin{document}

\narrowtext

{\large \bf Comment on ``Spin Polarization and Magnetic Circular Dichroism in
Photoemission from the $2p$ Core Level of Ferromagnetic Ni''
}\vspace{0.3cm}

Recently, Menchero\cite{Menchero} applied the 4-sites cluster
model\cite{Falicov} to the interpretation of the $2p$ spin-resolved
x-ray photoemission spectra (SRXPS) in Ni\cite{See95}. In this Comment
we show, by applying the Ni$_4$ cluster to the
$L_{2,3}$ magnetic circular dichroism (MCD),
that the Ni ground state is not well described by this model
and that it cannot provide a satisfactory description of all
magnetic dichroic experiments.

\begin{figure}
\epsfxsize 10.0cm
\inseps{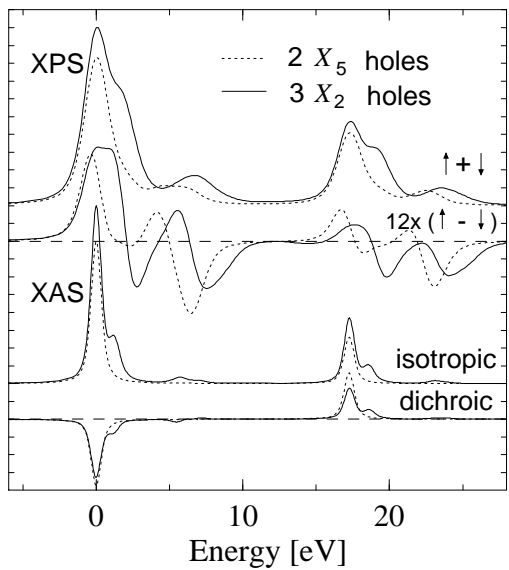}{0.8}
\caption{ The lower part shows the isotropic and circular dichroic
$L_{2,3}$ x-ray absorption spectra (XAS) and the upper part gives the sum
and the difference of the minority ($\uparrow$) and majority ($\downarrow$)
$2p$ XPS spectra, in the same geometry as in Ref.~\protect\cite{Menchero}.
Calculations are done for the Ni$_4$ cluster with two
$X_5$ (dotted) and three $X_2$ (solid) holes.  }
\label{clustersp:fig}
\end{figure}

The excitation of the core electron into the valence shell makes MCD more
sensitive to ground-state properties than SRXPS, where the difference
between the majority and minority spectrum is a result of final-state
interactions between the core hole and the polarized valence shell. The
calculation of the MCD spectrum, including a finite valence spin-orbit
coupling (Fig.~\ref{clustersp:fig}, dotted line), directly shows two major
discrepancies between theory and experiment.  First, the integrated
intensities at the two spin-orbit split edges do not correspond to the
experimentally observed ones.  From the relations of these intensities to
ground state expectation values of $L_z$ and $S_z$\cite{Thole92,Carra93} we
find $\left\langle L_{z}\right\rangle /\left\langle S_{z}\right\rangle
=0.35$. This should be compared with the experimental value of
$\left\langle L_{z}\right\rangle / \left\langle S_{z}\right\rangle
=0.19$\cite{Carra93,Chen}. It is clear that the Ni$_4$ cluster
overestimates the orbital magnetic moment.  Second, the satellite
structures, of mainly $d^{8}\longrightarrow \underline{p}d^{9}$ character,
are absent.

\noindent

These discrepancies are a direct result of the choice of the Ni$_4$
cluster. This model has a ground state consisting mainly of two holes of
$X_5$ ($t_{2g}$-like) symmetry. This preference for one particular
$k$-point leads to a ground state that overestimates the orbital moment and
has no $d^8$ character, since the two holes entirely avoid each
other\cite{Falicov}.  Choosing a two-hole ground state of $X_2^2$
($e_g$-like, which is the second lowest state) or mixed $X_2,X_5$ character
lowers the orbital moment to some extent, but still leads to a small $d^8$
character ($<$ 3 \%).

We were only able to obtain a ground state with a relatively small
$\left\langle L_{z}\right\rangle /\left\langle S_{z}\right\rangle$ of 0.24
and a significant $d^8$ character (12 \%) by having three holes of
predominantly $X_2$ character.  Unfortunately, the good agreement for SRXPS
is then lost, as Fig.~\ref{clustersp:fig} (solid line) shows.

In conclusion, although the Ni$_4$ cluster includes more information
regarding the Ni band structure with respect to the Anderson impurity
model\cite{Jo91,vdLaan92}, it also favors very peculiar ground states which
are incompatible with a coherent picture of all dichroism experiments.  In
many cases the less specific Anderson impurity model might even provide a
better description of the local ground state properties of metals. Any
attempts to improve the situation by increasing the cluster size would
imply formidable computational efforts.

\vspace{0.3cm}

\noindent
Nicola Manini, Michel van Veenendaal, and Massimo Altarelli\\
European Synchrotron Radiation Facility\\
B.P. 220\\
F-38043 Grenoble C\'edex - France


\widetext
\end{document}